\documentstyle[11pt,twoside,vlad,epsf]{article}

\heads{Relativistic Astrophysics and Cosmology}{BLACK HOLES IN PHASE TRANSITION \ \
{\rm M.Yu.~Khlopov, R.V.~Konoplich, S.G.~Rubin, 
A.S.~Sakharov}}

\begin{document}

\Arthead{1}{10}

\Title{FIRST ORDER PHASE TRANSITIONS AS A SOURCE OF BLACK HOLES IN
THE EARLY UNIVERSE\footnote{Invited talk on 10th Russian Gravitational Conference.} }%
{M.Yu.~Khlopov$^{1,2,3}$, R.V.~Konoplich$^{1,2}$, 
S.G.~Rubin$^{1,2}$, 
A.S.~Sakharov$^{1,2,4}$}%
{$^1$ Center for CosmoParticle Physics "Cosmion" \\
$^2$ Moscow Engineering Physics Institute \\
$^3$ Institute for Applied Mathematics \\
$^4$ Labor f$\ddot u$r H$\ddot o$chenergiephysik, ETH-H$\ddot o$nggerberg}

\Abstract{ A new mechanism of black hole formation in a first 
order phase 
transition is
proposed. In vacuum bubble collisions the interaction of bubble 
walls leads
to the formation of nontrivial vacuum configuration. The 
consequent collapse
of this vacuum configuration induces the black hole formation with 
high
probability. The primordial black holes that have been created by 
this way at the end of 
first order inflation could give essential contribution into the 
total density of 
the early Universe. The possibilities to establish some nontrivial 
restrictions on the inflation models with first order phase 
transition are discussed.  }

\section{Introduction}

At present time black holes (BH) can be created only by a 
gravitational
collapse of compact objects with mass more than about three Solar 
mass \cite{1}. However at the early stage of evolution of the 
Universe there 
where no limits on the mass of BH formed by several mechanisms. 
The simplest one is a collapse of strongly inhomogeneous regions 
just after the end of 
inflation 
(see e.g. \cite{2}). Another possible source of BH could be a 
collapse of 
cosmic
strings \cite{3} that are produced in early phase transitions with 
symmetry
breaking. The collisions of the bubble walls \cite{4,5} created at 
phase
transitions of the first order can lead to a primordial black hole 
(PBH)
formation.

We discuss here new mechanism of PBH production in the collision 
of two
vacuum bubbles. The known opinion of the BH absence in such 
processes is
based on strict conservation of the original O(2,1) symmetry. 
Whereas
there are ways to break it . Firstly, the radiation of scalar 
waves
indicates the entropy increasing and hence the permanent breaking 
of the
symmetry during the bubble collision. Secondly, the vacuum decay 
due to
thermal fluctuation does not possess this symmetry from the 
beginning. The
simplest example of a theory with bubble creation is a scalar 
field theory
with two non degenerated vacuum states. Being stable at a 
classical level,
the false vacuum state decays due to quantum effects, leading to a
nucleation of the bubbles of true vacuum and their subsequent 
expansion. \cite {6} The potential energy of the false vacuum is converted into a 
kinetic
energy of the bubble walls thus making them highly relativistic in 
a short
time. The bubble expands till it collides with another one. As it 
was shown
in \cite{4,5} a black hole may be created in the collision of 
several
bubbles. Our investigations show that BH can be created as well 
with a
probability of order unity in the collisions of only two bubbles. 
It
initiates the enormous production of BH that leads to essential 
cosmological
consequences discussed below.

In Section 2 the evolution of the field configuration in the 
collisions of
bubbles is discussed. The BH mass distribution is obtained in 
Section 3. In
Section 4 cosmological consequences of the BH production in bubble
collisions at the end of inflation are considered.

\section{Evolution of field configuration in collisions of vacuum
bubbles}

Consider a theory where a probability of false vacuum decay equals 
$\Gamma $
and difference of energy density between the false and true vacuum 
outside equals $\rho_V$. Initially bubbles are produced at rest 
however walls of the bubbles quickly increase their velocity up to 
the speed of light $v=c=1$ because a conversion of the false 
vacuum energy into its kinetic ones is energetically favorable.

Let us discuss dynamics of collision of two true vacuum bubbles 
that have
been nucleated in points $({\bf r}_1,t_1),({\bf r}_2,t_2)$ and 
which are
expanding into false vacuum. Following papers \cite{4,7} let us 
assume for
simplicity that the horizon size is much greater than the distance 
between
the bubbles. Just after collision mutual penetration of the walls 
up to the
distance comparable with its width is accompanied by a significant 
potential
energy increase \cite{8}. Then the walls reflect and accelerate 
backwards.
The space between them is filled by the field in the false vacuum 
state
converting the kinetic energy of the wall back to the energy of 
the false
vacuum state and slowdown the velocity of the walls. Meanwhile the 
outer
area of the false vacuum is absorbed by the outer wall, which 
expands and
accelerates outwards. Evidently, there is an instant when the 
central region
of the false vacuum is separated. Let us note this false vacuum 
bag (FVB) does 
not possess
spherical symmetry at the moment of its separation from outer 
walls but wall
tension restores the symmetry during the first oscillation of FVB. 
As it was 
shown 
in \cite{7}, the further evolution of FVB consists of several 
stages:

1) FVB grows up to the definite size $D_M$ until the kinetic 
energy of its wall becomes zero;

2) After this moment the false vacuum bag begins to shrink up to a 
minimal size $D^{*}$;

3) Secondary oscillation of the false vacuum bag occurs.

The process of periodical expansions and contractions leads to 
energy losses
of FVB in the form of quanta of scalar field. It has been shown in 
the \cite
{7,9} that only several oscillations take place. On the other 
hand,
important note is that the secondary oscillations might occur only 
if the
minimal size of the FVB would be larger than its gravitational 
radius, $%
D^{*}>r_g$. The opposite case ($D^{*}<r_g$ ) leads to the BH 
creation with
the mass about the mass of the FVB. As we will show later the 
probability of
BH formation is almost unity in a wide range of parameters of 
theories with
first order phase transitions.

\section{Gravitational collapse of FVB and BH creation}

Consider in more details the conditions of converting FVB into BH. 
The mass $M$ of FVB can be calculated in a framework of a specific 
theory 
and can be
estimated in a coordinate system $K^{\prime }$ where the colliding 
bubbles
are nucleated simultaneously. The radius of each bubble 
$b^{\prime}$ in
this system equals to half of their initial coordinate distance at 
first
moment of collision. Apparently the maximum size $D_M$ of the FVB 
is of the
same order as the size of the bubble, since this is the only 
parameter of
necessary dimension on such a scale: $D_M=2b^{\prime }C$. The 
parameter $C\simeq 1$
is obtained by numerical calculations in the framework of each 
theory, but its exact numerical value does not affect 
significantly conclusions.

One can find the mass of FVB that arises at the collision of two 
bubbles of
radius:

\begin{equation}
\label{one}M=\frac{4\pi }3\left( Cb^{\prime }\right) ^3\rho _V 
\end{equation}
This mass is contained in the shrinking area of false vacuum. 
Suppose for
estimations that the minimal size of FVB is of order wall width 
$\Delta $.
The BH is created if minimal size of FVB is smaller than its 
gravitational
radius. It means that at least at the condition

\begin{equation}
\label{two}\Delta <r_g=2GM 
\end{equation}
the FVB can be converted into BH (where G is the gravitational 
constant).

As an example consider a simple model with Lagrangian

\begin{equation}
\label{three}L=\frac 12\left( \partial _\mu \Phi \right) ^2-\frac 
\lambda
8\left( \Phi ^2-\Phi _0^2\right) ^2-\epsilon \Phi _0^3\left( \Phi 
+\Phi
_0\right) . 
\end{equation}
In the thin wall approximation the width of the bubble wall can be 
expressed
as $\Delta =2\left( \sqrt{\lambda }\Phi _0\right) ^{-1}$. Using 
(2) one can
easily derive that at least FVB with mass

\begin{equation}
\label{four}M>\frac 1{\sqrt{\lambda }\Phi _0G}
\end{equation}
should be converted into BH of mass M. The last condition is valid 
only in
case when FVB is completely contained in the cosmological horizon, 
namely $%
M_H>1/\sqrt{\lambda }\Phi _0G$ where the mass of the cosmological 
horizon at
the moment of phase transition is given by $M_H\cong 
m_{pl}^3/\Phi_{0}^2$. Thus 
for the potential (3) at
the condition $\lambda >(\Phi_0/m_pl)^2$ the BH is formed. This 
condition is 
valid for any realistic
set of parameters of theory.

The mass and velocity distribution
of FVBs, supposing its mass is large enough to satisfy the 
inequality (2), has been found in \cite{mi}. This distribution can 
be written in the terms of dimensionless mass 
$\mu \equiv \left( \frac \pi 3\Gamma \right)
^{1/4}\left( \frac M{C\rho _v}\right) ^{1/3}$:

\begin{equation}
\label{12}
\begin{array}{c}
\frac{dP}{\Gamma ^{-3/4}Vdvd\mu }=64\pi \left( \frac \pi 3\right) 
^{1/4}\mu
^3e^{\mu ^4}\gamma ^3J(\mu ,v), \\ J(\mu ,v)=\int_{\tau 
_{}}^\infty d\tau
e^{-\tau ^4},\tau _{-}=\mu \left[ 1+\gamma ^2\left( 1+v\right) 
\right] .
\end{array}
\end{equation}

The numerical integration of (\ref{12}) revealed that the 
distribution 
is rather narrow. For example the number of BH with mass 30 times 
greater than 
the average one is
suppressed by factor $10^5$. Average value of the non dimensional 
mass is equal 
to $\mu=0.32$. It allows to relate the average mass of BH and 
volume containing 
the BH at the moment of the phase transition:

\begin{equation}
\label{MV}\left\langle M_{BH}\right\rangle =\frac C4\mu ^3\rho
_v\left\langle V_{BH}\right\rangle \simeq 0.012\rho 
_v\left\langle
V_{BH}\right\rangle . 
\end{equation}

\section{First order phase transitions in the early Universe}

Inflation models ended by a first order phase transition hold a 
dignified
position in the modern cosmology of early Universe (see for 
example \cite{10,11}). The interest to these models is due to, 
that such models are able
to generate the observed large-scale voids as remnants of the 
primordial
bubbles for which the characteristic wavelengths are several tens 
of Mpc. 
\cite{11}. A detailed analysis of a first order phase transition 
in the
context of extended inflation can be found in \cite{12}. Hereafter 
we will
be interested only in a final stage of inflation when the phase 
transition
is completed. Remind that a first order phase transition is 
considered as
completed immediately after establishing of true vacuum 
percolation regime.
Such regime is established approximately when at least one bubble 
per unit
Hubble volume is nucleated. Accurate computation \cite{12} shows 
that first
order phase transition is successful if the following condition is 
valid:
\begin{equation}
\label{14}Q\equiv \frac{4\pi }9\left( \frac \Gamma {H^4}\right)
_{t_{end}}=1. 
\end{equation}
Here $\Gamma$ is the bubble nucleation rate. In the framework of 
first order
inflation models the filling of all space by true vacuum takes 
place due to
bubble collisions, nucleated at the final moment of exponential 
expansion.
The collisions between such bubbles occur when they have comoving 
spatial
dimension less or equal to the effective Hubble horizon 
$H_{end}^{-1}$ at
the transition epoch. If we take $H_0=100hKm/\sec /Mpc$ in 
$\Omega =1$
Universe the comoving size of these bubbles is approximately $%
10^{-21}h^{-1}Mpc$. In the standard approach it believes that such 
bubbles
are rapidly thermalized without leaving a trace in the 
distribution of
matter and radiation. However, in the previous section it has been 
shown
that for any realistic parameters of theory, the collision between 
only two
bubble leads to BH creation with the probability closely to 100\% 
. The mass of 
this BH is given by (see (\ref{MV}))
\begin{equation}
\label{15}M_{BH}=\gamma _1M_{bub} 
\end{equation}
where $\gamma _1\simeq 10^{-2}$ and $M_{bub}$ is the mass that 
could 
be
contained in the bubble volume at the epoch of collision in the 
condition of
a full thermalization of bubbles. The discovered mechanism leads 
to a new
direct possibility of PBH creation at the epoch of reheating in 
first order
inflation models. In standard picture PBHs are formed in the early 
Universe
if density perturbations are sufficiently large, and the 
probability of PBHs
formation from small post- inflation initial perturbations is 
suppressed
exponentially. Completely different situation takes place at final 
epoch of
first order inflation stage; namely collision between bubbles of 
Hubble size
in percolation regime leads to PBHs formation with masses

\begin{equation}
\label{16}M_0=\gamma _1M_{end}^{hor}= 
\frac{\gamma _1}2\frac{m_{pl}^2}{H_{end}}, 
\end{equation}
where $M_{end}^{hor}$ is the mass of Hubble horizon at the end of 
inflation. According to (\ref{MV}) the initial mass fraction of 
this 
PBHs is given by $\beta _0\approx\gamma _1/e\approx 6\cdot 10^{-
3}$.
For example, for typical value of 
$H_{end}\approx 4\cdot 10^{-6}m_{pl}$ 
the initial mass fraction $\beta $ is
contained in PBHs with mass $M_0\approx 1g$.

In general the Hawking evaporation of mini BHs could give rise to 
a variety possible end states.
It is generally assumed, that evaporation proceeds until the PBH 
vanishes
completely \cite{21}, but there are various arguments against this 
proposal 
(see e.g. \cite{22}). If one supposes that BH evaporation 
leaves a stable relic, then it is naturally to assume that it has  
a mass of order $m_{rel}=km_{pl}$, 
where $k\simeq 1\div 10^2$. We can investigate the consequences of 
PBH 
forming at the percolation epoch after first order inflation, 
supposing that the stable relic is a result of its evaporation.
As it follows from our above consideration the PBHs are 
preferentially formed with a typical mass $M_0$ at a single time 
$t_1$. Hence the total density $\rho$ at this time is
\begin{equation}
\label{totdens}
\rho (t_1)=\rho_{\gamma}(t_1)+\rho_{PBH}(t_1)=
\frac{3(1-\beta_0)}{32\pi t_1^2}m_{pl}^2+
\frac{3\beta_0}{32\pi t_1^2}m_{pl}^2
\end{equation}
The evaporation time scale can be written in the following form
\begin{equation}
\label{evop}
\tau_{BH}=\frac{M_0^3}{g_*m_{pl}^4}
\end{equation}
where $g_*$ is the number of effective massless degrees of 
freedom. 

Let us derive the density of PBH relics. There are two 
distinct possibilities to consider.

The Universe is still radiation dominated at $\tau_{BH}$. This 
situation will be hold if the following condition is valid 
$\rho_{BH}(\tau_{BH})<\rho_{\gamma}(\tau_{BH})$. It is possible to 
rewrite this condition in terms of Hubble constant at the end of 
inflation
\begin{equation}
\label{con1}
\frac{H_{end}}{m_{pl}}>\beta_0^{5/2}g_*^{-1/2}\simeq 10^{-6}
\end{equation}
Taking the present radiation density fraction of the Universe to 
be $\Omega_{\gamma_0}=2.5\cdot 10^{-5}h^{-2}$ ($h$ being the 
Hubble 
constant in the units of $100km\cdot s^{-1}Mpc^{-1}$), and using 
the standard values for the present time and time when the density 
of matter and radiation become equal, we find the contemporary 
densities fraction of relics
\begin{equation}
\label{reldens}
\Omega_{rel}\approx 10^{26}h^{-2}
k\left(\frac{H_{end}}{m_{pl}}\right)^{3/2}
\end{equation}
It is easily to see that relics overclose the Universe 
($\Omega_{rel}>>1$) for any reasonable $k$ and 
$H_{end}>10^{-6}m_{pl}$. 

The second case takes place if the Universe becomes PBHs dominated 
at period 
$t_1<t_2<\tau_{BH}$. This situation is realized under 
the condition  $\rho_{BH}(t_2)<\rho_{\gamma}(t_2)$, which can be 
rewritten in the form
\begin{equation}
\label{con2}
\frac{H_{end}}{m_{pl}}<10^{-6}.
\end{equation}
The present day relics density fraction takes the form
\begin{equation}
\label{reldens2}
\Omega_{rel}\approx 10^{28}h^{-2}
k\left(\frac{H_{end}}{m_{pl}}\right)^{3/2}
\end{equation}
Thus the Universe is not overclosed by relics only if the 
following condition is valid
\begin{equation}
\label{con3}
\frac{H_{end}}{m_{pl}}\le 2\cdot 10^{-19}h^{4/3}k^{-2/3}.
\end{equation}
This condition implies that the masses of PBHs created at the end 
of inflation have to be larger then
\begin{equation}
\label{massr}
M_0\ge 10^{11}g\cdot h^{-4/3}\cdot k^{2/3}.
\end{equation}
From the other hand there are a number of well--known cosmological 
and astrophysical limits \cite{15} which 
prohibit the creation of PBHs in the mass range (\ref{massr}) with 
initial fraction of mass density closed to 
$\beta_0\approx 10^{-2}$. 

So one have
to conclude that the effect of the false vacuum bag mechanism of 
PBH
formation makes impossible the coexistence of stable remnants of 
PBH
evaporation with the first order phase transitions at the end of 
inflation.

{\it Acknowledgements}.
This work was partially performed in the framework of Section 
"Cosmoparticle 
physics" of Russian State Scientific Technological Program 
"Astronomy. 
Fundamental Space Research", International project Astrodamus, 
Cosmion-ETHZ and 
Eurocos-AMS.


\small
\begin{thebibliography}{1} \itemsep=-5pt


\bibitem{1}  J. R. Oppenheimer, H. Sneider, Phys. Rev. {\bf 56}, 
455 (1939)

\bibitem{2}  A.G.Polnarev, M.Yu Khlopov, Sov. Phys. Usp. {\bf 
145}, 369 (1985)

\bibitem{3}  S. Hawking, Phys. Lett. {\bf B231}, 237 (1989);
A. Polnarev and R. Zembowicz, Phys. Rev. {\bf D43}, 1106 (1991)

\bibitem{4}  S.W.Howking, I.G.Moss, J.M.Stewart, Phys. Rev. {\bf 
D26}, 2681 (1982)

\bibitem{5}  I.G.Moss, Phys. Rev. {\bf D50}, 676 (1994)

\bibitem{6}  R.Watkins, M.Widrow, Nucl. Phys. {\bf B374}, 446 
(1992)

\bibitem{7}  S.Coleman, Phys.Rev. {\bf D15}, 2929 (1977)

\bibitem{8}  R.V.Konoplich, Sov. Nucl. {\bf 32}, 1132 (1980)

\bibitem{9}  T.I Belova, A.E.Kudryavtsev, Physica {\bf D32}, 18 
(1988)

\bibitem{mi} M.Yu.Khlopov, R.V.Konoplich, S.G.Rubin, A.S.Sakharov, 
hep-ph/9807343

\bibitem{10}  D.La, P.J.Steinhardt, Phys. Rev. Lett. {\bf 62}, 376 
(1989);
R.Holman, E.W.Kolb, Y.Wang, Phys. Rev. Lett. {\bf 65}, 17 (1990); 
R.Holman,
E.W.Kolb, S.Vadas, Y.Wang, Phys. Rev. {\bf D43}, 3833, (1991); 
F.C.Adams,
K.Freese, Phys. Rev. {\bf D43}, 353 (1991); E.J.Copeland, 
A.R.Liddle, D.H.Lyth,
E.D.Stewart, D.Wands, Phys. Rev. {\bf D49}, 6410 (1994)

\bibitem{11}  F.Occhionero, L.Amendola, Phys. Rev. {\bf D50}, 4846 
(1994);
L.Amendola, C.Baccigalupi, F.Occhionero, R. Konoplich, S.Rubin, 
Phys. Rev. 
{\bf D54}, 7199 (1996),

\bibitem{12}  M.S. Turner, E.J. Weinberg and L.M. Widrow, Phys. 
Rev. {\bf D46},
2384 (1992)

\bibitem{21}  S.W.Hawking, Nature (London) {\bf 248}, 30 (1974)

\bibitem{22}  M.A.Markov, Quantum Gravity, Proceeding of the 2nd 
Seminar,
Moscow, USSR, 1981, edited by M.A.Markov and P.C.West (Plenum, New 
York,
1984); Ya.B.Zeldovich, Quantum Gravity, Proceeding of the 
2nd
Seminar, Moscow, USSR, 1981, edited by M.A.Markov and P.C.West 
(Plenum, New
York, 1984); J.D.Barrow, E.J.Copeland, A.R.Liddle, Phys. Rev. 
{\bf D46}, 645
(1992); B.J.Carr, J.H.Gilbert, Phys. Rev. {\bf D50}, 4853 
(1994); S.O.Alexeyev, M.V.Pomazanov, Phys. Rev. {\bf D55}, 
2110 (1997); I.G.Dymnikova, Int. Jour. Mod. Phys. {\bf D5}, 529 
(1996)

\bibitem{15}  Ya.B.Zeldovich, A.A.Starobinski, JETP Letters, {\bf 
24}, 616 (1976); P.D.Naselski, Sov. Astron. J. Lett. {\bf 4}, 387 
(1978); S.Miyama, K.Sato, Prog. Theor. Phys. {\bf 59}, 1012 
(1978); D.Lindley, Mon. Not. R. Astron. Soc. {\bf 193}, 593 
(1980); Ya.B.Zeldovich, A.A.Starobinski, M.Yu.Khlopov,
V.M.Chechetkin,
Sov. Astron. J. Lett. {\bf 3}, 208 (1977); T.Rothman, R.Matzaner,
Astrophys. Space. Sci. {\bf 75}, 229 (1981); J.H. MacGibbon and 
B.J. Carr, 
Astrophys. J. {\bf 
371}, 447 (1991)


\end{thebibliography}
\end{document}